\documentclass[prd, twocolumn, superscriptaddress,altaffilletter,showpacs,nofootinbib]{revtex4}

\usepackage{amssymb}
\usepackage{amsfonts}
\usepackage{amsthm}
\usepackage{graphicx}
\usepackage{color}
\usepackage{amsmath}
\usepackage{latexsym}
\usepackage{makeidx}\makeindex
\usepackage{fancyhdr}
\usepackage[utf8]{inputenc} 
\usepackage[dvipsnames]{xcolor}
\usepackage{verbatim}

\newcommand{\be}{\begin{equation}}
\newcommand{\ee}{\end{equation}}
\newcommand{\bea}{\begin{eqnarray}}
\newcommand{\eea}{\end{eqnarray}}

\begin{document}


\title{A simple estimation of the size of the molecules using a pencil lead}

\author{Ricardo Medel Esquivel}
\email{rmedele1500@alumno.ipn.mx}
\author{Isidro Gómez Vargas}%
\affiliation{ 
Instituto Politécnico Nacional, CICATA-Legaria, Ciudad de México, CP 11500, México. 
}%
\author{J. Alberto Vázquez}
\affiliation{ 
Instituto de Ciencias F\'isicas, Universidad Nacional Aut\'onoma de M\'exico, Cuernavaca, Morelos, 62210, M\'exico 
}%
\author{Ricardo García Salcedo}
\affiliation{ 
Instituto Politécnico Nacional, CICATA-Legaria, Ciudad de México, CP 11500, México. 
}%
\date{\today}


\maketitle

%


    One of the main topics of elementary physics is the idea that every material is composed of "little particles that move around in perpetual motion, attracting each other when they are a little distance apart, but repelling upon being squeezed into one other"\cite{Feynman}. These particles could be atoms or molecules. Atoms are the smallest part into which any material can be divided. Whereas when several atoms are joined together, molecules are formed.
    
    Some interesting experiments to estimate the size of such atoms or molecules have been done that do not involve sophisticated equipment. One of these early experiments was conducted by Lord Rayleigh (1842-1919), which consisted of a small drop of oil spread to form a circular patch on the surface of the water. With a few simple calculations it is possible to determine the size of the oil molecule composition and therefore to provide an estimate of the diameter of the carbon atom\cite{Rayleigh, Farooq}.
    
    The main aim of this article is that students, at the basic level of education, gain a quantitative understanding of the size of molecules by performing a simple experiment easily designed within the classroom.
    All they need is a pencil lead, millimeter paper and a measuring instrument. Of course, we assume that all molecules are approximately the same size\cite{Swartz}.   
    
\begin{figure}[h]
        \includegraphics[scale=0.55]{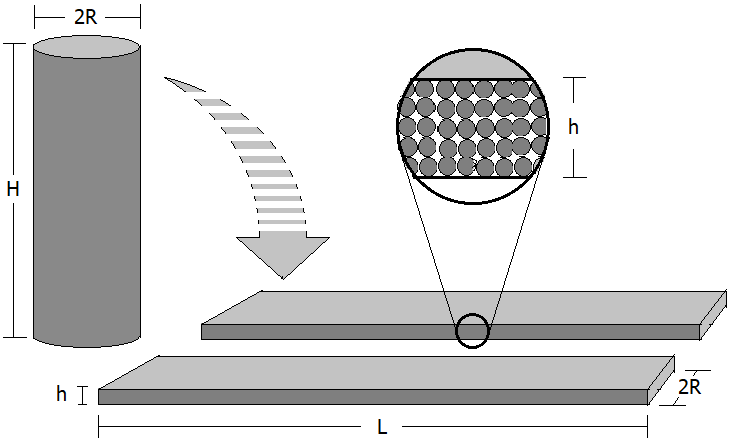}
        \caption{\label{Esquema} It is assumed that the volume of the graphite cylinder is distributed in $n$ identical boxes, whose height $h$ gives us an estimate of the maximum size a molecule could have. In detail, it is considered that $h$ is not the height of a single molecule but of many.}
        \end{figure}

    The pencil lead is composed of graphite molecules (the fourth most abundant chemical element in the Universe \cite{Hazen2013}), which we can imagine as identical spherical particles. These molecules form the macroscopic structure of the pencil, which has the shape of an elongated cylinder and hence its volume is given by:
        \begin{equation}
            V_C=\pi R^2 H,
        \end{equation}
    where $R$ is the radius of the pencil lead and $H$  its height as we can see in Figure \ref{Esquema}.

    If we draw a line on a sheet of millimeter paper, keeping the pencil straight, part of the pencil material will have moved to the paper matrix, forming a very thin parallelepiped or tiny height box, whose volume is given by:
        \begin{equation}
            V_B= 2RLh,
        \label{VB}    
        \end{equation}
    where $L$ is the length of the line, $2R$ is the width of the line and $h$ is the height of the box.
    
    While the volume of graphite spent after drawing the line is:
        \begin{equation}
            V_C'=\pi R^2 (H-H').
        \label{VC}
        \end{equation}

    Under this assumption the material is deposited entirely on the surface of the paper, without loss, then we can match the volume spent on the pencil lead would be equal to the volume deposited on the paper. See Figure \ref{Esquema}.
    
    And to make the effect more visible, it is possible to draw $n$ equal lines, of known length. Then, the following equality is satisfied $V_C'=V_B$, therefore from (\ref{VB}) and (\ref{VC}):
    
        \begin{equation}
            h=\pi \frac{R(H-H')}{2nL}.
        \end{equation}
    
    This height $h$ can be considered as an upper bound for the size of the graphite molecules, considered that is not the height of a single molecule but of many. Taking $n$ large enough to significantly reduce the length of the pencil lead is possible to calculate this estimate numerically.
    
    According to the manufacturer's specifications the HB pencil lead have $H = 60$ mm and $2R = 0.5$ mm (this is a popular standard measure, however there are other presentations with different length and diameter). 
        \begin{figure}[h]
        \includegraphics[width=7cm]{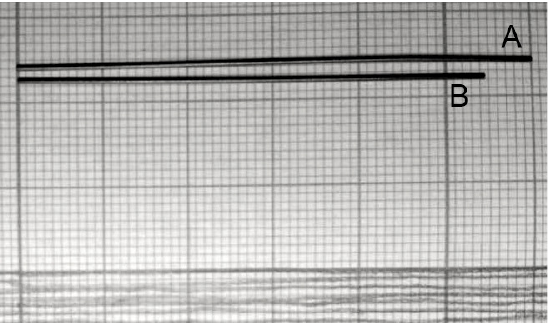}
        \caption{\label{fig:epsart} It is shown the "lead" (A) before drawing the lines and the "lead" (B) after draw some lines in the millimeter paper. The measure of what has been consumed of graphite is that it is used to calculate the volume.}
        \end{figure}        
    
    We found that for this experiment it is easier to draw short lines, about 10 cm long, in the central columns of the millimeter paper, starting from above. We performed the test with $n=50$ y $L=100$ mm (Fig. 2), and obtained that $H'=59.5$ mm, and therefore
         \begin{equation}
            h=\pi \frac{0.25mm(0.5mm)}{2(50)(100mm)}=0.000039 mm = 3.9 \times 10^{-8}m.
        \end{equation}   
    
    This result is reasonable as a higher level for the size of the molecules, the individual size of the so-called graphene \footnote{Graphene is a single layer of carbon atoms arranged in an hexagonal lattice, with one carbon atom at each vertex.} sheets has been systematically measured and varies from 2 to 20 nm ($2\times 10^{-9}m$ to $2 \times 10^{-8} m$) \cite{Shen2018}. The order of magnitude of the result does not change significantly when the number of lines drawn or their length is increased. Similar results can be found by performing analog experiments to the one presented here, although those are more sophisticated and require more equipment to carry them out \cite{Schaefer}.




\end{document}